\documentclass{appolb}
\usepackage{epsfig}

\newcommand{\be}{\begin{equation}}
\newcommand{\ee}{\end{equation}}
\newcommand{\ba}{\begin{array}}
\newcommand{\ea}{\end{array}}
\newcommand{\bea}{\begin{eqnarray}}
\newcommand{\eea}{\end{eqnarray}}

\begin{document}
\title{Pattern formation in a stochastic model of cancer growth
}
\author{
Anna Ochab-Marcinek
\address{Marian~Smoluchowski Institute of Physics,
Jagellonian University, Reymonta~4, PL-30--059~Krak\'ow, Poland}
}
\maketitle
\begin{abstract}

We investigate noise-induced pattern formation in a model of cancer growth based on
Michaelis-Menten kinetics, subject to additive and multiplicative noises. We analyse stability
properties of
the system and discuss the role of diffusion and noises in the system's dynamics. We find that random dichotomous
fluctuations in the immune response intensity along with Gaussian environmental noise
lead to emergence of a spatial pattern of two phases, in which cancer cells, or, respectively,
immune cells predominate.
\end{abstract}



\section{Introduction}
The study of population dynamics covers a wide range of fields such as ecology, cellular and molecular biology
and medicine (see e.g. \cite{Murray,Spagnolo_Valenti_Fiasconaro,Sachs_Hlatky,Hahnfeld_Sachs}).
The models of population growth based on nonlinear ordinary differential
equations, despite their simplicity, can often capture the essence of
complex biological interactions and explain characteristics of proliferation phenomena.
However, biological processes are not purely deterministic: systems existing
in nature are subject to natural noises.

The presence of noise in biological systems gives rise
to a rich variety of dynamical effects. Random fluctuations may
 be regarded not only as a mere source of disorder
 but also as a factor which introduces positive and organizing rather than disruptive changes
 in the system's dynamics. Some of the more important
 examples of noise induced effects are
 stochastic resonance \cite{Gammaitoni_Hanggi,Spagnolo_Valenti_Fiasconaro},
  resonant activation \cite{Doering_Gadoua,Iwaniszewski,Ochab_Gudowska},
  noise-induced transitions \cite{Lefever,Broeck_Parrondo},
noise-enhanced stability \cite{Spagnolo}
and pattern formation \cite{Murray,Spagnolo_Valenti_Fiasconaro,Valenti_Fiasconaro_Spagnolo}.

 The effect of cell-mediated immune surveillance against
 cancer \cite{GARAY,Stepanova,Vladar} may be a specific illustration of the coupling
  between noise and a biological system. Most of tumoral cells bear antigens which are recognised as
 strange by the immune system. A response against these antigens may be mediated either by immune
 cells such as T-lymphocytes or other cells, not directly related to the immune system (like
  macrophages or natural killer cells). The process of damage to tumour proceeds via infiltration
   of the latter by
  the specialised cells which subsequently develop a cytotoxic activity against the
  cancer cell-population.
  The series of reactions between the cytotoxic cells and the tumour tissue may be
  considered to be well approximated by a saturating, enzymatic-like process
  whose time evolution equations are similar to the standard Michaelis-Menten kinetics
  \cite{GARAY,Prigogine_Lefever}.
  Random variability of kinetic parameters defining that process may effect 
the extinction of the tumour.

In our previous article
\cite{Ochab_Gudowska} we discussed the noise-induced effect of resonant activation in a spatially
homogeneous model of cancer growth \cite{Prigogine_Lefever} based upon the above-mentioned kinetic scheme.
 In the present paper, we focus on the study of a spatially inhomogeneous system, namely, we
 investigate how global environmental noise as well as fluctuations in the immune response
 parameter effect the emergence of spatial patterns.


\section{The Model}

The interaction between cancer cells and cytotoxic cells
will be described by use of the predator-prey model
based upon the Michaelis-Menten kinetic scheme \cite{Ochab_Gudowska,Lefever,GARAY,
Prigogine_Lefever,Lefever_Horsthemke,Mombach}. This model is a classical one and has been
extensively studied since the 1970s. Its validity has been verified experimentally e.g. in
\cite{Garay_Lefever}, where the authors examined the mechanism of immune rejection of a tumour
induced by Moloney murine sarcoma virus. The behaviour of the cellular populations may be represented
by the following scheme:

 First, the cytotoxic
cells bind to the tumour cells at rate $k_{1}$; subsequently, the cancer cells which have been bound are
killed and the complex dissociates at rate $k_{2}$; finally, dead cancer cells decay at a rate $k_3$.
The process can be described schematically:
\be
X\ +\ Y\ \longrightarrow \!\!\!\!\!\!\!\! ^{k_{1}} \ \ \ \ Z\ \longrightarrow \!\!\!\!\!\!\!\! ^{k_{2}}\ \ \ \  Y\ +\ P\ \longrightarrow \!\!\!\!\!\!\!\! ^{k_{3}}\ \ \ \  Y.
\ee
$X$ represents here the population of tumour cells. $Y$, $Z$ and
$P$ are populations of active cytotoxic cells, bound cells and dead tumour cells,
 respectively.
 Following the original presentation \cite{Prigogine_Lefever}, we assume that
 (i) cancer cells undergo replication at a rate proportional
to a time constant $\lambda$; (ii) as a result of cellular replication
in limited volume, a diffusive propagation of cancer cells is possible, with
a transport coefficient proportional to the replication rate and local density of cancer cells;
 (iii) dead cancer cells undergo elimination at rate $k_{3}$; (iv)
   free cytotoxic cells can move with a ``diffusion'' coefficient $D$.
The spatio-temporal evolution of the tumour due to the above processes can
be then described by a set of balance equations:
\be\label{basic_system}
\left\{
\begin{array}{lll}
\frac{\partial x}{\partial t}&=&\lambda[1-(x+p)]x-k_{1}yx+ \lambda(1-p)\nabla^2 x +\lambda x\nabla^2 p \\

\frac{\partial y}{\partial t}&=&-k_{1}yx+k_{2}z+D\nabla^2 y \\
\frac{\partial z}{\partial t}&=&k_{1}yx-k_{2}z \\
\frac{\partial p}{\partial t}&=&k_{2}z-k_{3}p
\end{array}
\right.
\ee
The $x(\overrightarrow r, t) ,y(\overrightarrow r, t),z(\overrightarrow r, t)$ and $p(\overrightarrow r, t)$
are local densities of cells at point $\overrightarrow r$. Finally, we impose an additional condition on the model: the total number of active and bound
 cytotoxic cells should remain constant:
\be\label{eq:E}
y(t)+z(t)=const.=E.
\ee

Below, we analyse several versions of the model:
\begin{itemize}
\item{Bifurcation analysis, based on a bare kinetics model
without spatial diffusion (Sec. \ref{sec:stability_analysis})}
\item{Incorporation of Fickian diffusion terms in evolution equations
for $x(\overrightarrow r,t)$, $p(\overrightarrow r,t)$ respectively, which leads to a wavefront
solution (Sec. \ref{sec:diff})}
\item{We determine the effect of the dichotomous switching in the kinetic parameter $k_1$
and discuss its role also for the model system in which each of the kinetic equations
 is additionally driven by a Gaussian noise term of the same intensity (Sec. \ref{sec:dich_no_diff}, \ref{sec:dich_gauss_no_diff},
\ref{sec:gauss_diff})}
\item{At the last step of the model complexity, we analyse how the joint effect of diffusion,
additive Gaussian fluctuations and independent dichotomic switching in $k_1$ parameter affect
 the cancer cells population dynamics (Sec. \ref{sec:all}).}
\end{itemize}
\section{Simulation}\label{sec:simulation}

The basic aim of this work was to study the behaviour of the system (\ref{basic_system}) subject to
 additive and multiplicative noises:

\be\label{sim_system}
\left\{
\begin{array}{lll}
\frac{\partial x}{\partial t}&=&\lambda[1-(x+p)]x-(k_{1}+\eta(t))yx+\\
 & & +\lambda(1-p)\nabla^2 x +\lambda x\nabla^2 p + \sigma\xi(\overrightarrow r,t)\\
\frac{\partial y}{\partial t}&=&-(k_{1}+\eta(t))yx+k_{2}z+D\nabla^2 y  + \sigma\xi(\overrightarrow r,t)\\
\frac{\partial z}{\partial t}&=&(k_{1}+\eta(t))yx-k_{2}z  + \sigma\xi(\overrightarrow r,t)\\
\frac{\partial p}{\partial t}&=&k_{2}z-k_{3}p + \sigma\xi(\overrightarrow r,t)
\end{array}
\right.
\ee
The multiplicative dichotomous Markovian noise $\eta(t)=\pm\Delta$ with mean frequency
 $\gamma$ and autocorrelation $<\eta(t)\eta(t')> = \frac{\Delta^2}{2}e^{-2\gamma|t-t'|}$
models fluctuations in immune response. The additive Gaussian noise $\xi(\overrightarrow r,t)$
with autocorellation
$<\xi(\overrightarrow r,t)\xi(\overrightarrow r',t')> = \delta(\overrightarrow r-\overrightarrow r')
\delta(t-t')$ depicts external environmental fluctuations. We assume that
its intensity $\sigma$ is same for each variable of the system.
Both noises are assumed to be statistically independent: $<\xi(\overrightarrow r,t)\eta(s)>=0$.

\subsection{Numerics}

We have solved the stochastic differential equations (\ref{sim_system}) numerically,
 using the Euler scheme, on a $128 \times 128$ square lattice. According to the statistical properties of $\eta(t)$,
  the waiting time between two switchings was generated from the exponential distribution.

Since $x,y,z$ and $p$ are densities, their values never can be greater than $1$ or less than $0$.
Consequently, following boundary conditions have been imposed on the simulated system: if $x$, $y$,
$z$ or $p$ becomes less than $0$ or greater than $1$ at a given time step, we adjust its value to $0$
 or to $1$, respectively.

\subsection{Simulation Results} \label{sec:simulation_results}

We performed a simulation with the following values of parameters:

\be
\begin{array}{l}
\lambda = 0.5,\  D = 0.05,\  \sigma = 0.01,\  \Delta = 0.5,\\
 k_1 =1.75,\  k_2 = 0.1,\  k_3 = 0.1,\  \gamma = 0.01.
\end{array}
\ee \label{eq:initial}
$\gamma$ is the mean rate of switching in $\eta(t)$.
Initial conditions:
\be \label{eq:initial_point}
x(\overrightarrow r,0)=0,\  y(\overrightarrow r,0)=0.4,
\  z(\overrightarrow r,0)=0,\  p(\overrightarrow r,0)=0.
\ee
The values of parameters and initial conditions have been chosen so that we
could obtain a distinct pattern: The immune response rates $k_1+\Delta$ and $k_1-\Delta$, along
with $\lambda$ lead to two different types of stationary behaviour (see Sec. \ref{sec:gauss_diff}). The mean switching rate
is one or two orders of magnitude slower than other kinetic parameters, which gives the system a
possibility to approach the stationary states. The environmental noise intensity $\sigma$ has been
chosen in such a way that it allows the system to jump between both mentioned states. The noise is,
however, weak enough to let the system form a pattern. At the selected value of $D$, the pattern
has sufficiently distinct boundaries and is relatively stable (at higher values of $D$ it
would dissolve quickly, whereas at smaller $D$ it would form small "grains"). Parameters $k_2$ and $k_3$ have been chosen by
 a trial-and-error procedure: $k_2$ is responsible for
the dissociation of $z$ into $y+p$. If the dissociation rate is large, then the active immune cells $y$ are
being released faster and thus the immune response is more effective. The $k_3$ parameter determines the
rate at which dead cancer cells are eliminated. Since dead cells occupy the living space,
this parameter controls the effective replication rate of
cancer cells.

\begin{figure}[t]
\begin{center}
\epsfig{figure=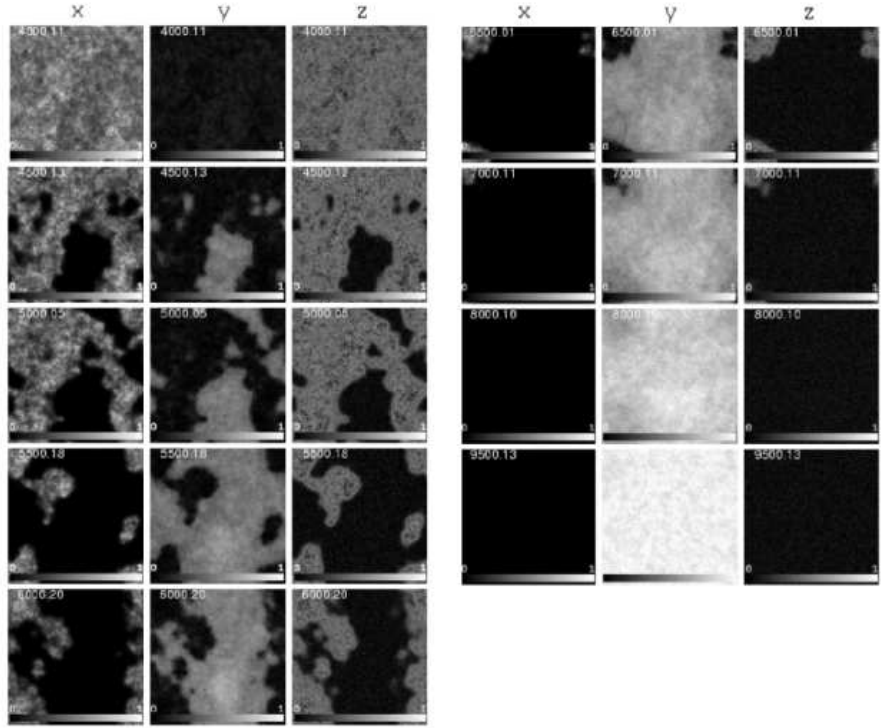, width=12cm}
\end{center}
\caption[]{\small Snapshots from the simulation of the system (\ref{sim_system}).
Time evolution of the spatial distribution
of $x$, $y$ and $z$. Time counter in upper left corner
of each image. Simulation time $T=10000$. Initial conditions: $x=0,y=0.4,z=0,p=0$ everywhere. Parameters:
$\lambda = 0.5, D = 0.05, \sigma = 0.01, \Delta = 0.5, k_1 =1.75, k_2 = 0.1, k_3 = 0.1,
\gamma = 0.01$. Light pixels: high concentration, dark pixels: low concentration of cells.
}
\label{fig:simulation}
\end{figure}

The simulation results are presented in Fig. \ref{fig:simulation}. After some time, we observe the emergence of the "$y$-phase" islands (where the immune cells
$y$ predominate) within the relatively homogeneous "$x$-phase" (in this phase cancer cells $x$
prevail). The phase boundaries move back and forth depending on the dichotomous changes in the
immune response intensity $k_1 \pm \Delta$.


\section{Analysis}

In order to explain the behaviour of the simulated system, we will analyse its stability
properties as well as the role of diffusion, additive noise and dichotomous noise in the $k_1$
 parameter.


\subsection{Stability analysis}\label{sec:stability_analysis}

The stationary points $\{ x^\star,y^\star,z^\star,p^\star\}$ of the system are given by:
\be\label{stationary_points_1}
\{0,y^\star,0,0\}
\ee
and
\be\label{stationary_points_2}
\left\{ \frac{k_3}{\lambda}\frac{\lambda-k_1 y^\star}{k_3+k_1 y^\star}, y^\star,
\frac{k_1 k_3 y^\star}{\lambda k_2}\frac{\lambda-k_1 y^\star}{k_3+k_1 y^\star},
\frac{k_1 y^\star}{\lambda}\frac{\lambda-k_1 y^\star}{k_3+k_1 y^\star} \right \}.
\ee

According to \ref{basic_system}, the value of $y^\star$ is arbitrary. However, if the condition
\ref{eq:E} is involved, it defines the value of $y^\star$:
\be\label{eq:E}
y^\star+z^\star=const.=E.
\ee

The sets of stationary points form two branches in the
 $x$-$y$-$z$-$p$-space.
 The branch (\ref{stationary_points_1}) changes its stability at the point
  $\{0,\frac{\lambda}{k_1},0,0\}$. It is repelling for
  $0 < y < \frac{\lambda}{k_1}$
  and attracting
  for $\frac{\lambda}{k_1} < y < 1$.
  The branch (\ref{stationary_points_2}) is attracting for
  $0 < y < \frac{k_3}{k_1}\left(-1+\sqrt{1+\frac{\lambda}{k_1}}\right)$.
  For $ \frac{k_3}{k_1}\left(-1+\sqrt{1+\frac{\lambda}{k_1}}\right) < y < 1$
   it consists of saddle points.

\begin{figure}[t]
\begin{center}
\epsfig{figure=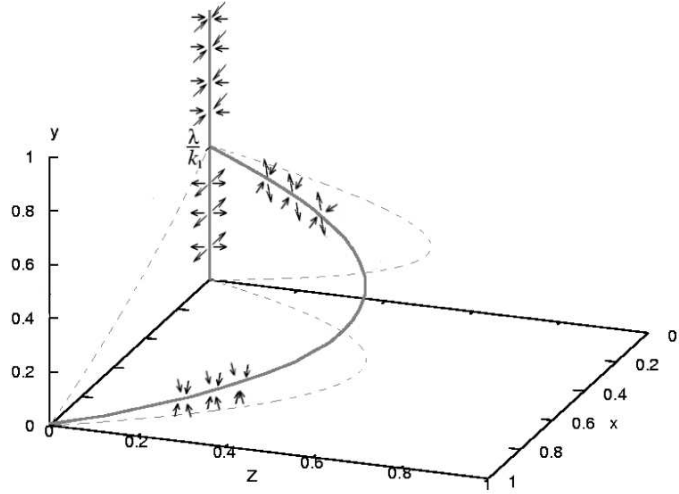, width=9cm}
\end{center}
\caption[]{\small Sets of stationary points (\ref{stationary_points_1}),
(\ref{stationary_points_2})
and their stability (shown only schematically by arrows), \textsl{x-y-z}-projection. Dashed lines:
planar projections of the curve. Parameters: $\lambda=0.5, k_1=1.25, k_2=0.1, k3=0.1$ }
\label{fig:branches}
\end{figure}

\begin{figure}[h]
\begin{center}
\epsfig{figure=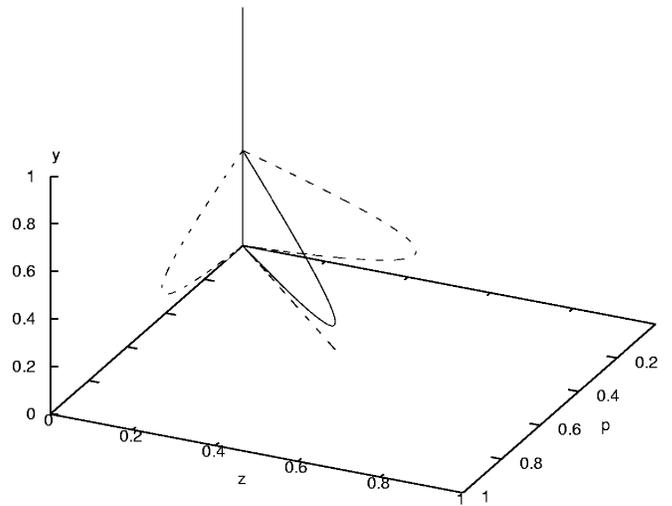, width=9cm}
\end{center}
\caption[]{\small Sets of stationary points (\ref{stationary_points_1}),(\ref{stationary_points_2})
in \textsl{y-z-p}-projection.
 Dashed lines: planar projections of the curve. Parameters: $\lambda=0.5, k_1=1.25, k_2=0.1, k3=0.1$ }
\label{fig:branches2}
\end{figure}

Trajectories of the deterministic system can lie only on the plane (\ref{eq:E}),
given by the initial conditions for $y$ and $z$. Hence, the stationary points of such
a system lie on the intersection of the plane (\ref{eq:E}) and the branches (\ref{stationary_points_1}),
(\ref{stationary_points_2}). Depending on the values of parameters and initial conditions,
 the system can have 1 (attracting), 2 (attracting
 and repelling) or 3 (attracting, saddle and attracting) stationary points (see Fig.
\ref{fig:branches}, \ref{fig:branches2}).

\begin{figure}[t]
\begin{center}
\epsfig{figure=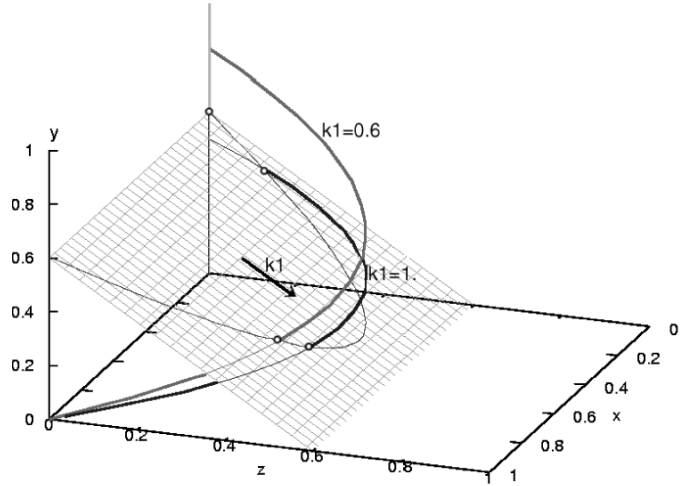, width=9cm}
\end{center}
\caption[]{\small Stationary points (open circles) lie on the intersection of the plane $y+z=E$
 and the curves given by (\ref{stationary_points_1}),(\ref{stationary_points_2}). As
$k_1$ grows, the number of intersections changes from 2 to 3 and, subsequently, from 3 to 1.
(An arrow shows schematically the direction in which the intersection points move as $k_1$ grows.)
The values of the remaining parameters are here:  $E=0.6, \lambda=0.6, k_2=0.1, k_3=0.1$}
\label{fig:fixed_parameters}
\end{figure}

When $E,k_2,k_3$ are fixed, the stability properties of the system depend
on the $k_1$ parameter (see Fig.
 \ref{fig:fixed_parameters}), i.e. the immune response efficiency, which, in our simulation,
 was controlled by the dichotomous noise.

In the next subsections, we will analyse how the presence of noises and diffusion
affects the behaviour of the system.


\subsection{Spatially inhomogeneous system without noise} \label{sec:diff}

The solutions to the deterministic system with diffusive terms have the form of travelling
 wavefronts \cite{Fife}.


\subsection{Spatially homogeneous system with dichotomous multiplicative noise} \label{sec:dich_no_diff}

\begin{figure}[h]
\begin{center}
\epsfig{figure=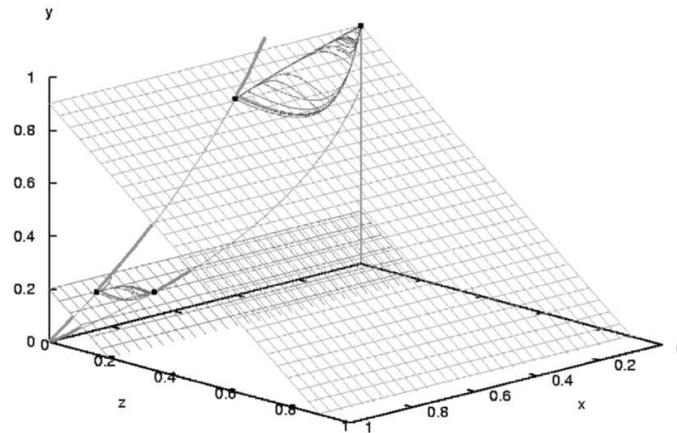, width=9cm}
\end{center}
\caption[]{\small Example trajectories of the system with dichotomous noise only, for two different
 values of $E=y+z$: $E=0.9$ and $E=0.2$. Parameters: $\lambda=0.5,
 \gamma=0.05,
 k_1=0.5, \Delta=0.25, k_2=1, k_3=1$. Simulation time $T=700$.}
 \label{fig:fig_eta_no_diff}
\end{figure}

After the introduction of the dichotomous noise $\eta(t)=\pm\Delta$ into the $k_1$ parameter,
 trajectories of the system lie on the planes $y+z=E$ and wander between two
 stationary points according to the current value of $k_1 \pm \Delta$ (see Fig. \ref{fig:fig_eta_no_diff}).

\newpage

\newpage

\newpage


\begin{figure}[t]
\begin{center}
\epsfig{figure=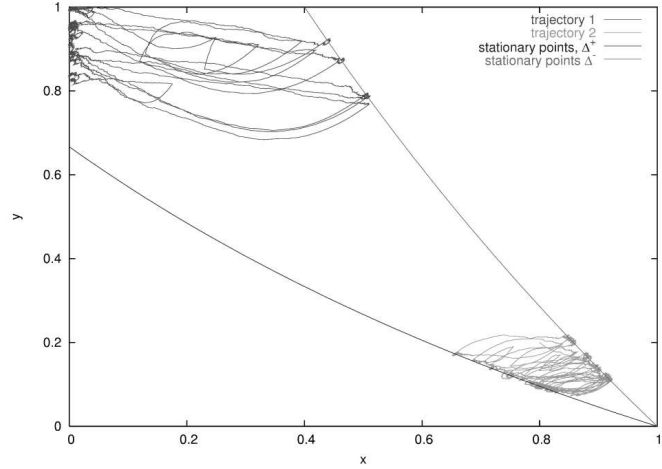, width=9cm}
\end{center}
\caption[]{\small Two example trajectories of the spatially homogeneous system with
multiplicative dichotomous noise, with additive Gaussian
 noise, starting from two different initial conditions: $y+z=E=0.9$
and $y+z=E=0.2$, in $x$-$y$-projection. $\sigma=0.002$, other parameters same as in
Fig. \ref{fig:fig_eta_no_diff}. To compare with the trajectories without additive noise, see
Fig. \ref{fig:fig_eta_no_diff}.
 }
\label{fig:fig_eta_xi_no_diff_1}
\end{figure}
\newpage


\subsection{Spatially homogeneous system with dichotomous multiplicative noise and additive
Gaussian white noise} \label{sec:dich_gauss_no_diff}

\begin{figure}[t]
\begin{center}
\epsfig{figure=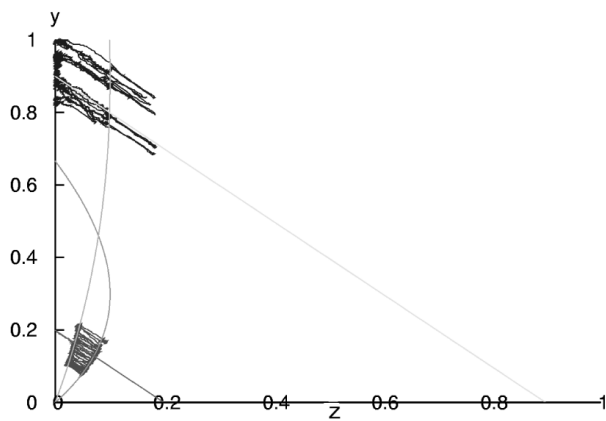, width=8cm}
\end{center}
\caption[]{\small Two example trajectories of the spatially homogeneous system with
multiplicative dichotomous noise and additive Gaussian noise,  starting with two different initial conditions: $E=0.9$
and $E=0.2$, in $y$-$z$-projection. The trajectories do not stay on their initial planes $E$
(denoted by sloped lines), but move up or down due to the Gaussian noise. Curves: branches of stationary points for switching value of
$k_1 \pm \Delta$
. $\sigma=0.002$, other parameters same as in
Fig. \ref{fig:fig_eta_no_diff}. }
\label{fig:fig_eta_xi_no_diff_2}
\end{figure}

In the next step of complexity, we add the term $\sigma\xi(\overrightarrow r,t)$ (Gaussian white noise) to each equation, assuming that
the additive noise acts in the same way on each variable of the system. Here, the trajectories do not
 stay on constant planes $y+z=E$ any more.
Due to the additive noise, they "slip off" from their initial planes
 (see Fig.\ref{fig:fig_eta_xi_no_diff_1},
\ref{fig:fig_eta_xi_no_diff_2}).


\subsection{System with Gaussian noise and diffusion} \label{sec:gauss_diff}
Let us examine a system with the additive Gaussian noise and diffusion, but without
the dichotomous noise. Instead, we will take into consideration
two situations where the intensity of the immune response is constant:
\be
K_1=k_1-\Delta
\ee
 and
\be
K_1=k_1+\Delta.
\ee
The value of $\Delta$ is same as in the first simulation (Sec. \ref{sec:simulation_results}), i.e.
it corresponds to one of the dichotomous noise states.

\begin{figure}[t]
\begin{center}
\epsfig{figure=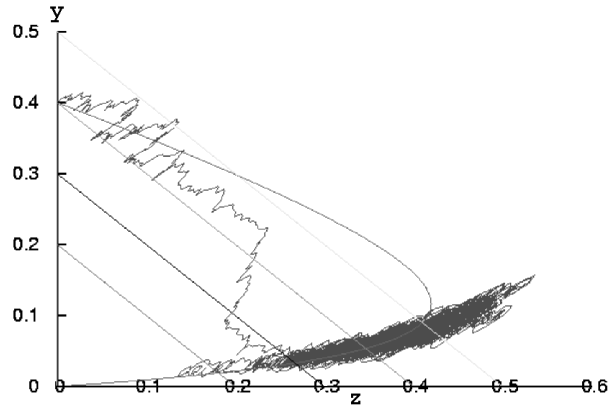, width=9cm}
\end{center}
\caption[]{\small An example trajectory of the system with additive Gaussian noise and
diffusion, recorded at point $[20,20]$ on the spatial lattice. The immune response parameter $K_1 = k_1 - \Delta
= 1.25$. The trajectory falls down onto the "$x$-phase" attractor. $\lambda=0.5, D=0.05, \sigma=0.01,
\Delta=0, k_2=0.1, k_3=0.1$. Initial conditions: $x=0, y=0.4, z=0, p=0$. Straight lines:
the profiles of example $E$ planes.}
\label{fig:k1-delta}
\end{figure}

\begin{figure}[t]
\begin{center}
\epsfig{figure=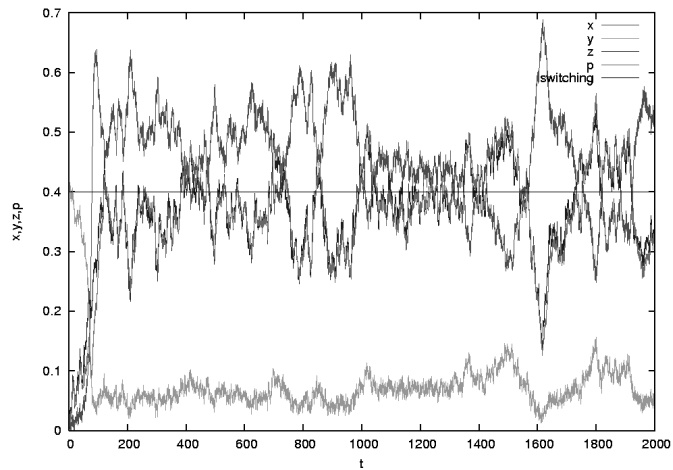, width=9cm}
\end{center}
\caption[]{\small Time-evolution of $x,y,z$ and $p$ in the system with additive Gaussian noise and
diffusion, recorded at point $[20,20]$ on the spatial lattice.  $K_1 = k_1 - \Delta = 1.25, \lambda=0.5, D=0.05, \sigma=0.01,
\Delta=0, k_2=0.1, k_3=0.1$. Initial conditions: $x=0, y=0.4, z=0, p=0$.}
 \label{fig:k1-delta_t}
\end{figure}

For $K_1=k_1-\Delta$, the "$x$-phase" is more stable. The trajectory starts with initial
 conditions (\ref{eq:initial_point}) which is exactly the point
$\{0,\frac{\lambda}{k_1},0,0\}$ where two branches of stationary points cross. It is then very likely
that the trajectory falls down onto the lower branch of the attractor and stays in its
neighbourhood (see Fig. \ref{fig:k1-delta},\ref{fig:k1-delta_t}).

\begin{figure}[t]
\begin{center}
\epsfig{figure=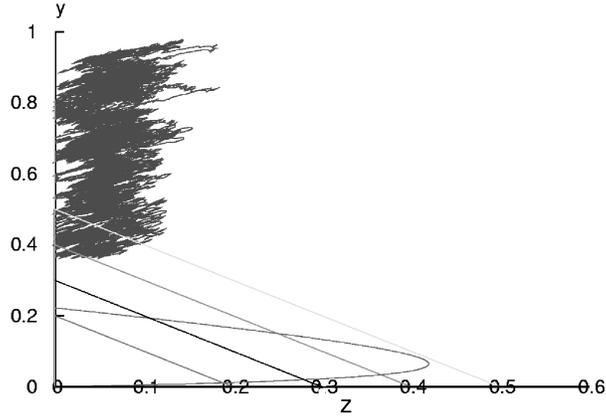, width=8cm}
\end{center}
\caption[]{\small An example trajectory of the system with additive Gaussian noise and
diffusion, recorded at point $[20,20]$ on the spatial lattice. The immune response parameter $K_1 = k_1 + \Delta
= 2.25$. The trajectory remains in the neighbourhood of the "$y$-phase" attractor and
climbs towards maximal values of $y$. $\lambda=0.5, D=0.05, \sigma=0.01,
\Delta=0, k_2=0.1, k_3=0.1$. Initial conditions: $x=0, y=0.4, z=0, p=0$ everywhere.}
\label{fig:k1+delta}
\end{figure}

\begin{figure}[h]
\begin{center}
\epsfig{figure=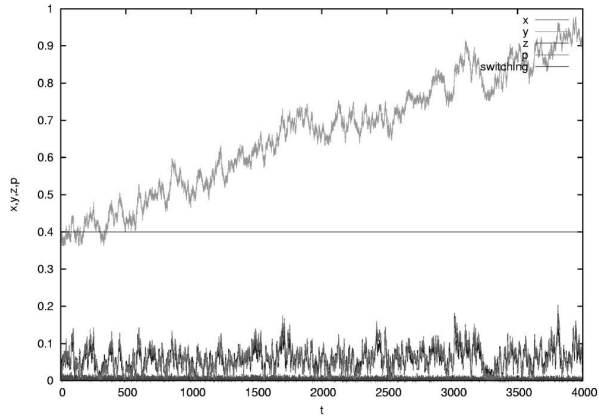, width=8cm}
\end{center}
\caption[]{\small The time-evolution of $x,y,z$ and $p$. in the system with additive Gaussian noise and
diffusion, recorded at point $[20,20]$ on the spatial lattice.  $K_1 = k_1 + \Delta = 2.25, \lambda=0.5, D=0.05, \sigma=0.01,
\Delta=0, k_2=0.1, k_3=0.1$. Initial conditions as above.}
 \label{fig:k1+delta_t}
\end{figure}

For $K_1=k_1+\Delta$, the "$y$-phase" is more stable. Starting far away from the lower branch of
the attractor, the trajectory remains close to the upper branch. Moreover, it climbs higher and higher
because of the reflecting boundary at $x=0, y=0, z=0, p=0 , x=1,y=1,z=1 ,p=1$
(see Fig. \ref{fig:k1+delta},\ref{fig:k1+delta_t}).
This condition imposed on
the boundaries is justified by the requirement that the population density cannot be negative nor
greater than $1$.

\begin{figure}[h]
\begin{center}
\epsfig{figure=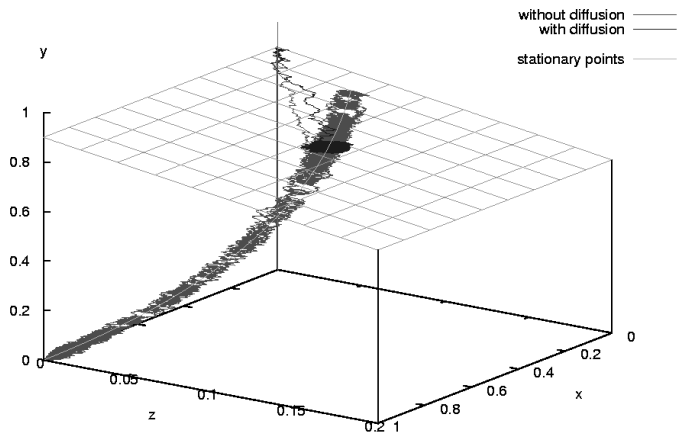, width=9cm}
\end{center}
\caption[]{\small Stabilising effect of diffusion. Gray: trajectory of a system with diffusion
($D=0.5$) and additive Gaussian noise, recorded at point $[20,20]$ on the spatial lattice. Black: trajectory of a system with additive Gaussian noise,
but without diffusion ($D=0$). Parameters: $\lambda=0.5, \sigma=0.005, k_1=0.25, k_2=1, k_3=1$.
Simulation time $T=3000$.}
\label{fig:fig_stab_diff}
\end{figure}

One can observe a synchronisation effect: the values of $y$, $z$ and $p$ decrease when $x$
increase, and {\em vice versa\/} (see Fig. \ref{fig:k1-delta_t},\ref{fig:k1+delta_t}). One can also notice the stabilising effect of diffusion. Fig. \ref{fig:fig_stab_diff} compares
trajectories of two systems: with and without diffusion, driven by additive Gaussian noise.
Without diffusion, the trajectory wanders up and down along the branch of stationary points.
The trajectory stabilised by diffusion stays longer in the neighbourhood of its initial plane $E$.

\subsection{System with Gaussian noise, dichotomous noise and diffusion} \label{sec:all}

\begin{figure}[h]
\begin{center}
\epsfig{figure=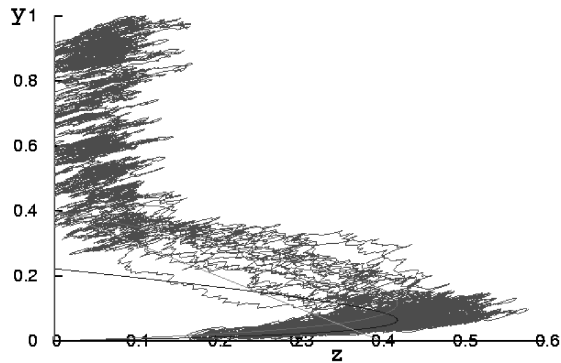, width=8cm}
\end{center}
\caption[]{\small Trajectory of the system (\ref{sim_system}) recorded at point $[20,20]$ on the spatial lattice, $y$-$z$-projection, simulation time
$T=10000$. The trajectory jumps between two stable branches of stationary points, but, finally,
it climbs up the "$y$-phase" branch. Initial conditions: $x=0,y=0.4,z=0,
p=0$ everywhere. Parameters:
$\lambda = 0.5, D = 0.05, \sigma = 0.01, \Delta = 0.5, k_1 =1.75, k_2 = 0.1,
 k_3 = 0.1, \gamma = 0.01$. Curves show how the shape of branches of stationary points
 changes with switching value of $k_1 \pm \Delta$. Sloped line: the initial plane $E=0.4$. }
\label{fig:simulation_traj}
\end{figure}

\begin{figure}[h]
\begin{center}
\epsfig{figure=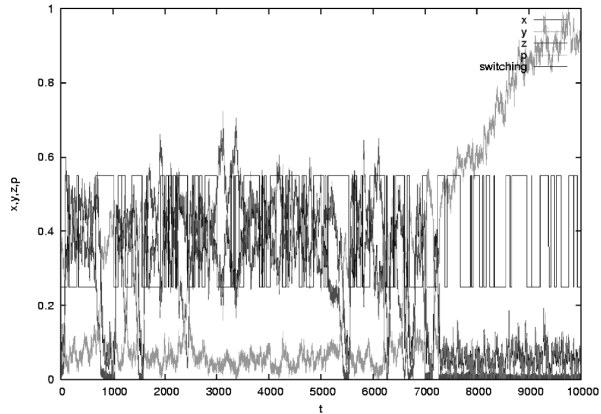, width=8cm}
\end{center}
\caption[]{\small Time evolution of $x,y,z$ and $p$ in system (\ref{sim_system}).
Simulation time $T=10000$. Initial conditions: $x=0,y=0.4,z=0,
p=0$ everywhere. Parameters:
$\lambda = 0.5, D = 0.05, \sigma = 0.01, \Delta = 0.5, k_1 =1.75, k_2 = 0.1,
 k_3 = 0.1, \gamma = 0.01$}
 \label{fig:simulation_traj_t}
\end{figure}

The system (\ref{sim_system}) is a combination of all cases analysed above. Dichotomous noise
 switches the system between states where either the "$x$-phase" or the "$y$-phase" is
 preferred. This causes the emergence of separate "islands"
 of these two phases. Their boundaries move due
 to diffusion and the direction and speed of the motion depends on the current value of $k_1+\eta(t)$
 (see Fig. \ref{fig:simulation},\ref{fig:simulation_traj_t}). In the \textit{x-y-z}-space, the trajectories climb up towards the region where $y$ is close to $1$
(the upper part of the $y$-phase attractor). This
"climbing" effect is caused by the boundary conditions imposed on the system:
Since the same positive or negative value of $\sigma\xi(\overrightarrow r,t)$ is being added to each equation,
 there is a certain preferred direction in which the the trajectory moves.
 It cannot, however, cross the boundaries and thus, near the attracting branch of
 (\ref{stationary_points_1}), the average direction of motion is "upwards"
 (i.e. towards the increasing values of $y$) because the motion in the opposite direction
 is blocked due to the boundary conditions (see Fig \ref{fig:simulation_traj}).

When trajectories get to the upper part of the "$y$-phase" attractor, they remain in its neighbourhood
for all time because the "$x$-phase" attractor is too distant. This distance is determined by the choice of parameters, namely by the position of point
$\{0,\frac{\lambda}{k_1},0,0\}$
where the other branch of stationary points begins (see Fig. \ref{fig:simulation_traj}).
This effect is the reason why the "$y$-phase" finally spreads all over accessible space in our simulation.

\newpage


\newpage



\section{Conclusions}

We have performed a simulation of a spatially inhomogeneous model of cancer growth with additive
 Gaussian noise and multiplicative dichotomous noise.
The multiplicative noise controls the efficiency of the immune response whereas the external
 environmental fluctuations have been modelled by the additive Gaussian noise.

The presence of noise in biological systems may be regarded not only as a mere source of disorder
 but also as a factor which introduces positive and organising rather than disruptive changes
 in the system's dynamics: In our model, we find that the presence of a global (i.e. depending
 only on time) multiplicative dichotomous
  noise in a system perturbed by a spatially inhomogeneous additive Gaussian noise leads to
   emergence of a spatial pattern of two "phases". The "phases" are distinct areas in which cancer cells,
   or, respectively, immune cells predominate. The pattern is not stable: domain boundaries move
 due to a diffusion effect, and the direction (and speed) of that motion is determined by the
 current value of the dichotomous noise, i.e. by the effective intensity of the immune response.

The spatial pattern emerges when the environmental noise intensity $\sigma$ is properly tuned.
 Combined with the multiplicative noise, it should allow the system to perform transitions between two "phases".
  Too strong additive noise would however dominate the picture, preventing the formation of a
  pattern. Additionally, the diffusion parameter $D$ should be carefully chosen, so that
 the pattern could have sufficiently distinct boundaries and be relatively stable.
If the diffusion rate is high, the pattern dissolves quickly, whereas
at small diffusion rate it only forms small "grains".The quantitative analysis of the interplay
between the above-mentioned factors in the process of pattern formation \cite{spagnolo_appb}
 merits a further study.

After a sufficiently long time, the immune cells prevail globally. This turns out to be the effect
of reflecting boundary conditions, which prevent the population densities from exceeding $0$ or
 $1$. The existence of such boundaries causes that the trajectories of the system prefer to move
in the direction of greater population of immune cells. A replacement of the additive Gaussian
noise $\xi(\overrightarrow r,t)$ with a multiplicative Gaussian noise, and a comparison with
the model described here, would be another interesting issue for future research.\\

The author would like to thank Dr. Ewa Gudowska-Nowak for helpful discussions and comments.


\newpage


\newpage


\begin{figure}[t]
\begin{center}
\epsfig{figure=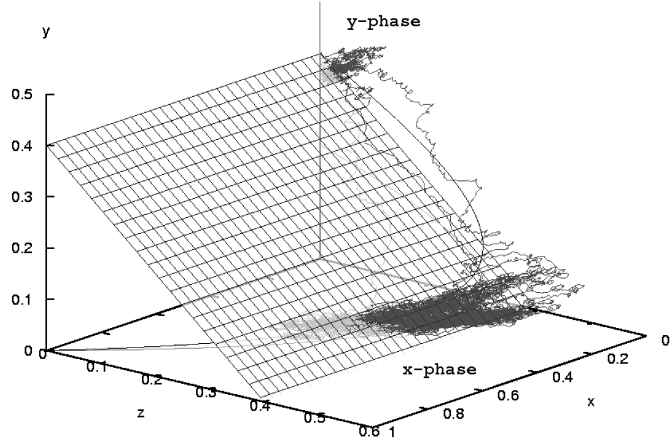, width=9cm}
\end{center}
\caption[]{\small Trajectory of the system (\ref{sim_system}) recorded at point
$[20,20]$ on the spatial lattice. Simulation time $T=4000$. The trajectory jumps between two possible phases. Initial conditions: $x=0,y=0.4,z=0,
p=0$ everywhere. Parameters:
$\lambda = 0.5, D = 0.05, \sigma = 0.01, \Delta = 0.5, k_1 =1.75, k_2 = 0.1,
 k_3 = 0.1, \gamma = 0.01$
}
\label{fig:two_phases}
\end{figure}

\newpage



\begin{thebibliography}{99}

\bibitem{Murray} J. D. Murray: {\em Mathematical Biology\/} (Springer-Verlag, Berlin, 1993)

\bibitem{Spagnolo_Valenti_Fiasconaro}  B. Spagnolo, D. Valenti, A. Fiasconaro, {\em Noise in Ecosystems: A Short Review\/},
Mathematical Biosciences and Engineering 1, in press (2004), arXiv:q-bio.PE/0403004

\bibitem{Sachs_Hlatky} R. K. Sachs, L. R. Hlatky, P. Hahnfeld, Math. Comp. Modelling 33 (2001) 1297.

\bibitem{Hahnfeld_Sachs} P. Hahnfeld, R. K. Sachs, In: {\em Advances in Mathematical Population Dynamics: Molecules, Cells and Man \/}, Edited by O.Arino, D.Axelrod, M.Kimmel. World Scientific Publishing Company, 1998.

\bibitem{Gammaitoni_Hanggi} L. Gammaitoni, P. H\"anggi, P. Jung, F. Marchesoni, Rev. Mod. Phys.70, 223 (1998)

\bibitem{Doering_Gadoua} C. R. Doering, J. C. Gadoua, Phys. Rev. Lett. 69, 2318 (1992)

\bibitem{Iwaniszewski} J. Iwaniszewski, Phys. Rev. E 68, 027105 (2003)

\bibitem{Ochab_Gudowska} A. Ochab-Marcinek, E. Gudowska-Nowak, Physica A 343 (2004) 557-572

\bibitem{Lefever} W. Horsthemke, R. Lefever: {\em Noise-Induced Transitions. Theory and Applications in Physics, Chemistry and Biology\/} (Springer-Verlag, Berlin, 1984)

\bibitem{Broeck_Parrondo} C. Van den Broeck, J. M. R. Parrondo, R. Torral, Phys. Rev. Lett. 73, 3395 (1994)

\bibitem{Spagnolo} B. Spagnolo, A. A. Dubkov, N. V. Agudov, Eur. Phys. J. B 40, 273-281 (2004)

\bibitem{Valenti_Fiasconaro_Spagnolo} D. Valenti, A. Fiasconaro, B. Spagnolo, Acta Physica Polonica B, Vol. 35, No. 4, April 2004

\bibitem{GARAY} R. P. Garay and R. Lefever, J. Theor. Biol. 73 (1978)  417.

\bibitem{Stepanova} N. Stepanova, Biophysics 24, 917-923 (1980)

\bibitem{Vladar} H. P. Vladar, J. A. Gonzalez, J. Theor. Biol. 227 (2004) 335-348

\bibitem{Prigogine_Lefever} I. Prigogine, R. Lefever, Comp. Biochem. Physiol. 67B (1980) 389.

\bibitem{Lefever_Horsthemke}
R. Lefever and W. Horsthemke,
{\em Bistability in fluctuating environments. Implications in tumor biology},
{\em Bulletin of mathematical biology} {\bf 41}, 469 (1978)


\bibitem{Mombach}
J. C. M. Mombach, N. Lemke, B. E. J. Bodmann and M. A. P. Idiart
{\em A mean-field theory of cellular growth},
{\em Europhysics letters} {\bf 56}(6), 923 (2002)

\bibitem{Garay_Lefever}
R. Garay and R. Lefever,
{\em Local description of immune tumor rejection},
{\em Developments in cell biology} {\bf 2}, {\em Biomathematics and cell kinetics} (1978)


\bibitem{Fife} P. C. Fife, J. Chem. Phys. Vol. 64, No. 2 (1976)

\bibitem{spagnolo_appb}
A. Fiasconaro, D. Valenti and B. Spagnolo,
{\em Nonmonotonic behavior of spatiotemporal pattern formation in a noisy Lotka-Volterra system},
{\em Acta Physica Polonica B} {\bf 35},4 (2004) 1491



\end{thebibliography}
\end{document}